# Transient nutations decay in diluted paramagnetic solids: a radiation damping mechanism

N. Ya. Asadullina, T. Ya. Asadullin

Department of General Physics,
Kazan national research technical university – KAI
10, K.Marx Str., Kazan, 420111, RUSSIA

A theory of the intensity and concentration dependent damping of nutation signals observed by Boscaino *et al.* (Phys. Rev B **48,** 7077 (1993); Phys. Rev. A **59,** 4087 (1999)) and by others in various two-level spin systems is proposed. It is shown that in diluted paramagnetic solids contribution of dipole-dipole interaction to the nutation decay is negligibly small. We elaborated a cavity loss (radiation damping) mechanism that explains the intensity- and concentration dependence of the damping. It is shown that instead of ordinary Bloch's transverse $T_2$ and longitudinal $T_1$ damping parameters the decay of transverse and longitudinal spin components in nutation process are described by one and the same intensity-, concentration-, frequency- and time dependent damping parameter.

## 1. Introduction

Quantum information processing includes excitation of a system of two-level particles (electron or nuclear spins – qubits) by a sequence of pulses of the resonant coherent electromagnetic field of various duration and intensity. The inserted to the system coherent information decays with some rate both during the pulse and after it (or between the pulses in the case of a sequence of pulses). This decay during the pulse with length of the order of a characteristic relaxation time manifests itself in the phenomenon of the transient nutations.

Transient nutations are the response of a system, initially at thermal equilibrium, to the abrupt application of a very intense resonant field in going to a new stationary state. In the case of the strong field, applied to the paramagnetic system, the response represents damped in time oscillations (Rabi oscillations) with the Rabi frequency $\chi = \gamma B_1$, where $B_1$ is the amplitude of the resonant field, $\gamma$ is gyromagnetic ratio. This effect, initially observed in the system of nuclear spins [1], later was seen also in the system of electron spins [2], in the optical transitions of atomic and molecular systems [3] etc.

In dilute paramagnetic solids, to be considered in this paper, there are several sources of damping of nutations. Here, the resonance line is strongly inhomogeneously broadened. As a result the signal decays due to *decoherence* of the precessing spins. Another source is a relaxation associated with the cavity losses and losses in the spin system. The losses in the spin system include spin-spin (dipole-dipole), spin-lattice, hyperfine and other possible interactions. Under the assumption that these interactions are described by the Bloch equations with constant parameters $T_1$ and $T_2$ (in solids usually $T_1 \gg T_2$), the amplitude of oscillations $V(t)$ is damped as [1]

$$V(t) \sim J_0(\chi t) K(t), \tag{1}$$

where $J_0(\chi t)$ is the zeroth order Bessel function, $K(t) = exp(-\Gamma_B t)$ is the decay function and $\Gamma_B = 1/(2T_2)$ is the decay rate in the Bloch model.

In contrast to this theoretical prediction, the decay observed in glassy silica containing $E'$ centers and in quartz crystals with $[AlO_4]^0$ centers [4,5] is faster than expected and depends on the driving-field intensity and on the concentration of the centers. The power dependence of the decay rate $\Gamma$ is well described by a linear dependence of $\Gamma$ on the induced Rabi frequency:

$$\Gamma = \alpha + \beta\chi, \qquad (2)$$

where $\alpha$ is equal to $\Gamma_B$ within the experimental uncertainties and $\beta$ is a parameter of the order of $10^{-2}$.

Furthermore, parameter $\beta$ increases on increasing spin concentration $n$. Later the decay rates of the form (2) were found in the other systems of the electron spins [6-12].

In a recent paper, Baibekov [13] found a contribution of the many-particle dipole-dipole interactions between the paramagnetic centers to the decay rate $\Gamma$, containing the dependence both on the amplitude of the microwave field (i.e. on frequency $\chi$) and on the concentration $n$ of paramagnetic centers (see also Ref. 14). However, obtained numerical value of $\beta$ is occurred to be significantly less than experimental one. By numerical solving the Schroedinger equation, the contributions of dipole-dipole interactions and of some other possible factors to the process of nutation damping were examined also in paper [11]. It is pertinent to note that papers [4, 5] are the sole publications where due to two-quantum excitation the transient nutations are observed and measured directly during the excitation pulse, whereas usually such measurements are performed indirectly by use of free induction decay and/or the echoes.

Experiments on transient nutations shed light on the physics of the interaction of the resonant field with the system of the two level particles and the nature of the relaxation processes in the system. In this context the results of the papers [4, 5, 6-12] yet remain largely unexplained. The purpose of the present article is to further study of the relevant relaxation processes. Below, it will be shown that the contribution of the dipole-dipole interactions to the damping of transient processes (including nutations) is negligibly small and that under intense resonant field the main contribution to these processes has its origin in the cavity losses or, more precisely, in the radiation damping [15]. The radiation damping is a consequence of the Joule heat at the cavity walls caused by precessing active spins and is a function of the concentration of the spins.

The paper is organized as follows. In Subsection 2.1. we obtain expressions for spectral and total concentrations of spins excited during the pulse solving a simple rate equation for a generic spin packet of the inhomogeneously broadened spin system. Since the spins are excited from an equilibrium state where the transverse components of spins are absent, the excitation process is described with use of the low intensity (equilibrium) relaxation parameters $T_1$ and $T_2$ known from experiment. Subsection 2.2. is devoted to solution of the Heisenberg equations for the excited spins; the equations account for both the cavity (radiation) losses and losses due to spin-spin interactions. As a result, general expressions are obtained for the nutation response and for intensity dependent relaxation parameters. Section 3 is devoted to numerical analysis of these general expressions. The results of previous sections and their relation to some early published results are briefly discussed in Section 4. The main results of the paper are summarized in Section 5.

## 2. Theory
### 2.1. Kinetics of excitation

The electron spin resonance lines in solids are inhomogeneously broadened, i.e. the resonance frequencies $\omega$ are distributed around the mean frequency $\omega_0$ according to a some law. Following Refs. 4, 5, we approximate the line form of the system of spins $S = ½$ under consideration by a Gaussian

$$f(\Delta) = (2\pi)^{-1/2} \sigma^{-1} \exp(-\Delta^2/2\sigma^2), \qquad \Delta = \omega - \omega_0 \qquad (3)$$

with a standard deviation $\sigma$. For comparison purposes, in Section IV two other distribution forms, a Lorentzian and a rectangular, will be considered.

The system is excited by resonant field $2H_1 \cos\omega_0 t$ tuned at the centre of the inhomogeneously broadened line. The absorption line of a narrow band of spins (spin packets) with detuning $\Delta = \omega - \omega_0$ is homogeneously broadened and at low intensities of the resonant field is described by a Lorentzian

$$g_L(\Delta) = \frac{1}{\pi} \frac{\Gamma_0}{\Delta^2 + \Gamma_0^2}, \qquad (4)$$

with the width $\Gamma_0 = 1/T_2$ and $T_2$ is the low power value of the transverse relaxation time. With increasing intensity the absorption (and the concentration of excited spin packets) undergoes a power dependent saturation (see below).

The concentration dependence of the response is of great importance and this is determined by two reasons. Firstly, the response is obviously proportional to the concentration $n_{ac}$ of the *active* spins excited during the pulse. Secondly, the response depends on the concentrations of the active and of the passive spins due to the concentration dependence of the relaxation mechanisms under consideration: the dipole interactions of the active spins with each other and with the passive spins are different because the active spins have transverse components but passive spins have the *z*-component only. Hence, their contributions to the decay process depend on the $n_{ac}$ and $n_{pas}$ respectively. As noted above and will be shown below, the radiation damping is a function of concentration of the active spins as well.

Since in nutation experiments the pulse length is of the order or shorter than the phase relaxation time, the excitation of spins is an intensity and time dependent transient process. Moreover, due to the inhomogeneous broadening the concentration of excited spins is also frequency dependent and as a result we are in need of spectral function $n_{ac}(\chi, \Delta, t)$ for generic spin packet $S_\alpha(\Delta, t)$ ($\alpha = x, y, z$). To our knowledge, the concentration dependence of the nutation response was taken into account for the first time in Ref. 16. Here this question is considered in somewhat more detail.

The system of the spins in equilibrium is characterized by populations $n_{10}$ of the lower and $n_{20}$ of the upper levels and with $n_0 = n_{10} + n_{20}$ being the full concentration of the paramagnetic centers. During the pulse, the time dependence of the spectral concentration of the spins at the upper level $n_2(\Delta, t)$ is described by the rate equation ($n_1(\Delta, t)$ is the concentration of the spins at the lower level)

$$\dot{n}_2(\Delta, t) = -(w_{21} + W_{21}) n_2(\Delta, t) + (w_{12} + W_{12}) n_1(\Delta, t). \qquad (5)$$

As it follows from experiments [4, 5] and will be shown below, the longitudinal and transverse relaxation parameters of the spin system undergo power-, concentration-, and time- dependent drastic changes. These changes are connected with the time-varying transverse components of the spins. Since the excitation process takes place from equilibrium state with no transverse components, the kinetics $n_2(\Delta, t)$ should be governed by the equilibrium values of the spontaneous transition probability $w_{21} + w_{12} = 1/T_1$ and of the probabilities of the induced transitions $W_{12} = W_{21}$. Since the absorption profile of the homogeneously broadened line is Lorentzian, the induced transition probability $W_{ij}$ follows the Lorentzian line and we can introduce the frequency-dependent spectral saturation parameter

$$s(\Delta) = s_0 \frac{\Gamma_0^2}{\Delta^2 + \Gamma_0^2}, \qquad s_0 = \chi^2 T_2 T_1 \qquad (6)$$

with $T_1$ and $T_2$ being the low-power (equilibrium) constant parameters [4, 5].

Because of large intensity of the resonant field in experiments [4, 5], its variation due to absorption by the spin system can be ignored and $\chi$, $W_{ij} \sim \chi$ are considered as constant parameters.

Spectral concentrations of the active $n_{ac}(\Delta,t)$ and of the passive $n_{pas}(\Delta,t)$ spins at the time $t$ are

$$n_{ac}(\Delta,t) = [n_2(\Delta,t) - n_{20}(\Delta)] f(\Delta) = \frac{\Delta n_0}{2} \frac{s(\Delta)}{1+s(\Delta)} [1 - \exp(-\Gamma_n(\chi)t)] f(\Delta), \quad (7)$$

$$n_{pas}(\Delta,t) = \Delta n_0 f(\Delta) - n_{ac}(\Delta,t),$$

$$\Delta n_0 = n_{10} - n_{20} = n_0 - 2n_{20} = n_0 \tanh \frac{\Delta E}{2kT}.$$

Here $\Delta n_0$ is the initial population difference, where $\Delta E = \hbar \omega_0$ is the energy difference between the upper and the lower levels, $k$ the Boltzmann constant, $T$ temperature and

$$\Gamma_n(\chi) = [1 + s(\Delta)] \frac{1}{T_1}. \quad (8)$$

The full concentration of the active and passive spins at time $t$ will be respectively

$$n_{ac}(t) = \int_{-\infty}^{\infty} n_{ac}(\Delta,t) d\Delta, \quad (9)$$

$$n_{pas}(t) = \Delta n_0 - n_{ac}(t). \quad (10)$$

*2.2. Heisenberg equations*

The nutation response is proportional to the ensemble average of the transverse spin component of the system $\langle M_y(t) \rangle = \sum_i \langle S_y^i(t) \rangle$. It is convenient to perform calculations of the response using the Heisenberg equations for spin operators. For i-th spin-packet of the inhomogeneously broadened system we have

$$\frac{dS_\alpha^i}{dt} = \frac{\partial S_\alpha^i}{\partial t} + (i/\hbar)[H^i, S_\alpha^i], \quad \alpha = x, y, z. \quad (11)$$

In the applied magnetic field $\mathbf{B}_0 \parallel \mathbf{z}$ and microwave field $2\mathbf{B}_1 \cos\omega_0 t \parallel \mathbf{x}$ (seen by spins due to two-quantum excitation) the Hamiltonian $H^i$ consists of

$$H^i = H_0^i + H_1^i + H_{SS}^i + H_R^i, \quad (12)$$

where first two terms on the right hand side describe interactions of the $S^i$ with the applied and microwave magnetic fields, respectively. To show that contribution from magnetic dipole interactions to the damping is small, we include in equation (12) term $H_{SS}^i$ describing interaction with the active spins. Term $H_R^i$ describes interaction with cavity feedback field $\mathbf{B}_R$. The feedback field is connected with current $i_y$ induced at the cavity walls by precessing spins. Note that equation (12) not includes the dipole interaction with the passive spins (the final result for interaction with the passive spins will be given below without calculations).

In the reference frame rotating with frequency $\omega_0$ around $z$ axis of the laboratory reference frame, one has

$$H_0^i = \hbar \Delta_i S_z^i; \quad H_1^i = \hbar \chi S_x^i;$$

$$H_{SS}^i = \sum_j u_{ij} [S_z^i S_z^j - \tfrac{1}{4}(S_+^i S_-^j + S_-^i S_+^j)]. \quad (13)$$

Here $u_{ij} = \dfrac{\gamma^2 \hbar^2}{r_{ij}^3}\left(1 - 3\cos^2\theta\right)$ is the coordinate function in the dipole interaction and in $H_{SS}^i$ we have omitted oscillating time-dependent terms. In describing the cavity feedback field $\mathbf{B}_R$ we proceed from the Kirchhoff law [15]. In resonance conditions of experiments [4, 5] the only contribution to current $i_y$ is due to the precessing spins, so $B_R \sim i_y \sim dM_y/dt$, $M_y = \gamma\hbar \sum_j S_y^j$ and, as a consequence, the radiation damping Hamiltonian for i-th spin is

$$H_R^i = \hbar\gamma B_R^i S_x^i = S_x^i \sum_j a_{ij} \frac{dS_y^j}{dt}. \tag{14}$$

The feedback field component $B_R^{ij}$ at the site of i-th spin due to j-th spin is a coordinate function, that we specify by $a_{ij} = a_0/r_{ij}^3$, where $a_0$ is a cavity geometrical factor. Hamiltonian $H_R^i$ is simplified as follows. The Heisenberg equations in the rotating reference frame are

$$\frac{dS_\alpha^i}{dt} = (i/\hbar)\left[H^i, S_\alpha^i\right]. \tag{15}$$

If we retain in the Hamiltonian $H^i$ only $H_0^i$ and $H_1^i$, the equation for $S_y^i$ will be

$$\frac{dS_y^i}{dt} = \Delta_i S_x^i - \chi S_z^i. \tag{16}$$

Substitution into equation (14) for $dS_y^j/dt$ gives final expression for the radiation damping Hamiltonian

$$H_R^i = S_x^i \sum_j a_{ij}\left(\Delta_j S_x^j - \chi S_z^j\right). \tag{17}$$

Using Hamiltonian $H^i$ (12) given by expressions (13) and (17), the Heisenberg equations for $S_\alpha^i$ are (the primed sum indicate $j \neq i$)

$$\frac{dS_x^i}{dt} = -\Delta_i S_y^i - \sum_j{}' u_{ij}\left(S_y^i S_z^j + \tfrac{1}{2} S_z^i S_y^j\right),$$

$$\frac{dS_y^i}{dt} = \Delta_i S_x^i - \chi S_z^i + \sum_j{}' u_{ij}\left(S_x^i S_z^j + \tfrac{1}{2} S_z^i S_x^j\right) - S_z^i \sum_j{}' a_{ij}\left(\Delta_j S_x^j - \chi S_z^j\right), \tag{18}$$

$$\frac{dS_z^i}{dt} = \chi S_y^i + \tfrac{1}{2}\sum_j{}' u_{ij}\left(S_x^i S_y^j - S_y^i S_x^j\right) + S_y^i \sum_j{}' a_{ij}\left(\Delta_j S_x^j - \chi S_z^j\right).$$

Summand with $a_{ii}$ in the radiation damping Hamiltonian gives negligibly small contributions to the driving force and to frequency shift. It is easier to solve system (18) in the reference frame transformed as follows:

$$\tilde{S}_x^i = \frac{1}{\Omega_i}\left(\chi S_x^i + \Delta_i S_z^i\right), \qquad \tilde{S}_y^i = S_y^i,$$

$$\tilde{S}_z^i = \frac{1}{\Omega_i}\left(\chi S_z^i - \Delta_i S_x^i\right), \tag{19}$$

$$S_x^i = \frac{1}{\Omega_i}\left(\chi \tilde{S}_x^i - \Delta_i \tilde{S}_z^i\right), \qquad S_z^i = \frac{1}{\Omega_i}\left(\chi \tilde{S}_z^i - \Delta_i \tilde{S}_x^i\right),$$

where $\Omega_i = \left(\Delta_i^2 + \chi^2\right)^{1/2}$ is the effective Rabi frequency. The new system of equations is

$$\frac{d\tilde{S}_x^i}{dt} = -S_y^i \frac{1}{\Omega_i} \times \left[\chi \sum_j{}' u_{ij} \frac{1}{\Omega_j}\left(\Delta_j \tilde{S}_x^j + \chi \tilde{S}_z^j\right) - \tfrac{1}{2}\Delta_i \sum_j{}' u_{ij} \frac{1}{\Omega_j}\left(\chi \tilde{S}_x^j - \Delta_j \tilde{S}_z^j\right)\right] - \tfrac{1}{2}\tilde{S}_z^i \sum_j{}' u_{ij} S_y^j + \frac{\Delta_i}{\chi} S_y^i \xi_i^R,$$

$$\frac{dS_y^i}{dt} = -\left(\Omega_i + \xi_i\right)\tilde{S}_z^i + \frac{\Delta_i}{2\Omega_i}\tilde{S}_x^i \sum_j{}' u_{ij}\frac{1}{\Omega_j}\left(\chi\tilde{S}_x^j - \Delta_j\tilde{S}_z^j\right) + \frac{1}{\Omega_i}\chi\tilde{S}_x^i\sum_j{}' u_{ij}\frac{1}{\Omega_j}\left(\Delta_j\tilde{S}_x^j + \chi\tilde{S}_z^j\right), \tag{20}$$

$$\frac{d\tilde{S}_z^i}{dt} = \left(\Omega_i + \xi_i\right)S_y^i + \frac{1}{2}\tilde{S}_x^i\sum_j{}' u_{ij}S_y^j.$$

Here $\xi_i$ is contribution to the effective Rabi frequency from the nonlinear mechanisms:

$$\xi_i = \xi_i^S + \xi_i^R,$$

$$\xi_i^S = \frac{1}{\Omega_i}\sum_j{}' u_{ij}\frac{1}{\Omega_j}\times\left[\left(\Delta_i\Delta_j - \frac{\chi^2}{2}\right)\tilde{S}_x^j + \left(\Delta_i + \frac{\Delta_j}{2}\right)\chi\tilde{S}_z^j\right] = \frac{1}{\Omega_i}\sum_j{}' u_{ij}\left(\Delta_i S_z^j - \frac{\chi}{2}S_x^j\right), \tag{21}$$

$$\xi_i^R = -\frac{\chi}{\Omega_i}\sum_j{}' a_{ij}\Omega_j\tilde{S}_z^j = \frac{\chi}{\Omega_i}\sum_j{}' a_{ij}\left(\Delta_j S_x^j - \chi S_z^j\right).$$

If we account for also interaction with the passive spins (denoted as $I$), its contribution to $\xi_i$ is

$$\xi_i^I = \frac{\Delta_i}{\Omega_i}\sum_j u_{ij}I_z^j. \tag{22}$$

It follows from experiments [4, 5] that nutation response represents well defined oscillations (the Rabi oscillations). Furthermore, it will be shown by numerical calculations below that contributions of the dipole interactions to the decay are negligibly small. Hence, equations (20) are simplified as

$$\frac{d\tilde{S}_x^i}{dt} = \frac{\Delta_i}{\chi}\xi_i^R S_y^i, \quad \frac{dS_y^i}{dt} = -\left(\Omega_i + \xi_i\right)\tilde{S}_z^i,$$

$$\frac{d\tilde{S}_z^i}{dt} = \left(\Omega_i + \xi_i\right)S_y^i. \tag{23}$$

Frequency parameters $\xi_i^S$ and $\xi_i^R$ are slowly varying in time characteristics of the local magnetic field at i-th site due to the dipolar interaction and the feedback field, respectively and are considered in equations (23) as time-independent. Each j-th component of $\xi_i^{S,R}$, however, has the oscillatory dependence on time. Parameter $\xi_i^I$ is really time-independent. The solutions to equations for $S_y^i$ and $\tilde{S}_z^i$ are

$$S_y^i(t) = S_y^i(0)\cos\left(\Omega_i + \xi_i\right)t - \tilde{S}_z^i(0)\sin\left(\Omega_i + \xi_i\right)t, \tag{24}$$

$$\tilde{S}_z^i(t) = \tilde{S}_z^i(0)\cos\left(\Omega_i + \xi_i\right)t + S_y^i(0)\sin\left(\Omega_i + \xi_i\right)t.$$

$\tilde{S}_x^i(t)$ that enters into parameter $\xi_i^S$ can be obtained from the first of equations (23) without the right hand side there. As a result, with the use of equations (19) one has

$$S_x^i(t) = \frac{\Delta_i\chi}{\Omega_i^2}S_z^i(0)\left[1 - \cos\left(\Omega_i + \xi_i\right)t\right],$$

$$S_y^i(t) = -\frac{\chi}{\Omega_i}S_z^i(0)\sin\left(\Omega_i + \xi_i\right)t, \tag{25}$$

$$S_z^i(t) = \frac{1}{\Omega_i^2}S_z^i(0)\left\{\Delta_i^2 + \chi^2\left[\cos\left(\Omega_i + \xi_i\right)t - 1\right]\right\}$$

in the rotating reference frame where we retain terms proportional to $S_z^i(0)$ only. Note that for $\xi_i = 0$ equations (25) represent the relaxation-free solution of the Bloch equations ($T_1 = T_2 = \infty$). It is seen from equations (21) and (25) that $\xi_i^{S,R}$ and $S_\alpha^i(t)$ are interrelated. Substitution of $S_\alpha^j(t)$ from equations (25) into definitions (21) of $\xi_i^S(t)$ and $\xi_i^R(t)$ gives

$$\xi_i^S = \frac{1}{\Omega_i} {\sum_j}' u_{ij} \frac{1}{\Omega_j^2} S_z^j(0) \times \left[ \left( \Delta_i \Delta_j - \frac{\chi^2}{2} \right) \Delta_j + \left( \Delta_i + \frac{\Delta_j}{2} \right) \chi^2 \cos(\Omega_j + \xi_j) t \right], \tag{26}$$

$$\xi_i^R = -\frac{\chi^2}{\Omega_i} {\sum_j}' a_{ij} S_z^j(0) \cos(\Omega_j + \xi_j) t.$$

Let us consider the average value of $S_y^i(t)$:

$$\langle S_y^i(t) \rangle = Tr\{\rho(t) S_y^i(t)\} = -Tr\left\{ \rho(t) \frac{\chi}{\Omega_i} S_z^i(0) \sin(\Omega_i + \xi_i) t \right\}, \tag{27}$$

where the equilibrium density matrix is

$$\rho(t) = \exp\left( -\sum_i \mu_i B_0 / kT \right) / Z \approx \frac{1}{2^{n_{ac}}} \prod_i \left( 1 - \frac{\hbar \omega_0 S_z^i(0)}{kT} \right) \frac{1}{2^{n_{pas}}} \prod_j \left( 1 - \frac{\hbar \omega_0 I_z^j}{kT} \right), \tag{28}$$

and where concentrations of the active spins $n_{ac}$ and the passive spins $n_{pas}$ are given by equations (9) and (10).

Thus, in the rotating reference frame one has

$$\langle S_y^i(t) \rangle = -\frac{1}{2^{(n_{ac}+1)}} \frac{1}{2^{n_{pas}}} \frac{\chi \hbar \omega_0}{kT} \frac{1}{\Omega_i} \times \sum_{\{m_l^S\}} \sum_{\{m_k^I\}} \prod_k \left\langle m_k^I \left| \prod_l \left\langle m_l^S \left| \sin(\Omega_i + \xi_i^S + \xi_i^I + \xi_i^R) t \prod_l \left| m_l^S \right\rangle \prod_k \left| m_k^I \right\rangle \right.\right.\right.\right., \tag{29}$$

where $\{m_l^S\}'$ and $\{m_k^I\}$ denote the configurations of $n_{ac} - 1$ active and $n_{pas}$ passive spins, respectively with $m = \pm 1$. Performing summation over all possible spin configurations, we obtain

$$\langle S_y^i(t) \rangle = -\frac{\hbar \omega_0}{4kT} \frac{\chi}{\Omega_i} \times \sin \Omega_i t \prod_{j \neq i} \cos\left[ (\alpha_{ij} + \gamma_{ij}) t \right] \prod_j \cos(\beta_{ij} t); \tag{30}$$

$$\alpha_{ij} = \alpha_{ij}^0 u_{ij},$$

$$\alpha_{ij}^0 = \frac{1}{2\Omega_i \Omega_j^2} \times \left[ \Delta_j \left( \Delta_i \Delta_j - \frac{\chi^2}{2} \right) + \chi^2 \left( \Delta_i + \frac{\Delta_j}{2} \right) \overline{\cos\left[ (\Omega_j + \xi_j) t \right]} \right];$$

$$\beta_{ij} = \beta_{ij}^0 u_{ij}, \qquad \beta_{ij}^0 = \frac{\Delta_i}{2\Omega_i}; \tag{31}$$

$$\gamma_{ij} = -a_0 \gamma_{ij}^0 \frac{1}{r_{ij}^3}, \qquad \gamma_{ij}^0 = -\frac{\chi^2}{2\Omega_i} \overline{\cos\left[ (\Omega_j + \xi_j) t \right]}.$$

Factors $\prod_{j \neq i} \cos\left[ (\alpha_{ij} + \gamma_{ij}) t \right]$ and $\prod_j \cos(\beta_{ij} t)$ should be averaged over positions $\mathbf{r_j}$ and resonance frequencies $\Delta_j$ of spins in the sample [17, 13]. Let us suppose that there is one relaxation mechanism only, say, dipole interaction $H_{SS}^i$ between active spins. Then one has [13]

$$\prod_{j \neq i} \langle \cos(\alpha_{ij} t) \rangle_{r_j, \Delta_j} = \exp\left\{ -n_{ac}(t) \int d\Delta_j f(\Delta_j) \int d^3 r_{ij} \left[ 1 - \cos(\alpha_{ij} t) \right] \right\} = \exp\left[ -\Gamma_S(\Delta_i, t) t \right],$$

$$\Gamma_S(\Delta_i, t) = C n_{ac}(t) \int d\Delta_j f(\Delta_j) |\alpha_{ij}^0|, \tag{32}$$

where $C = 4\pi^2 g^2 \mu_B^2 / (9\sqrt{3}\hbar)$ (here $g$ is g-factor, $\mu_B$ Bohr magneton). Similar calculations for the interaction with passive spins, $H_{SI}^i$, give

$$\prod_j \langle \cos(\beta_{ij} t) \rangle_{r_j} = \exp[-\Gamma_I(\Delta_i, t) t],$$

$$\Gamma_I(\Delta_i, t) = C n_{pas}(t) |\beta_{ij}^0|, \tag{33}$$

and for radiation damping

$$\prod_{j \neq i} \overline{\cos(\gamma_{ij} t)}_{r_j, \Delta_j} = \exp\left\{-n_{ac}(t) \int d\Delta_j f(\Delta_j) \int d^3 r_{ij} [1 - \cos(\gamma_{ij} t)]\right\} = \exp[-\Gamma_R(\Delta_i, t) t],$$

$$\Gamma_R(\Delta_i, t) = a_1 n_{ac}(t) \int d\Delta_j f(\Delta_j) |\gamma_{ij}^0|, \tag{34}$$

where $a_1 \sim a_0$ is another numerical parameter. The same averaging process applied to $\overline{\cos[(\Omega_j + \xi_j) t]}$ in $\alpha_{ij}^0$ and $\gamma_{ij}^0$ (in equations (31)) leads to $\overline{\cos[(\Omega_j + \xi_j) t]} = \cos(\Omega_j t) \exp[-\Gamma_{S,R}(\Delta_j, t) t]$, respectively.

Below, it will be shown by numerical calculations that decay parameters $\Gamma_S(\Delta_i, t)$ and $\Gamma_I(\Delta_i, t)$ are negligibly small and that main contribution to the nutations decay is due to the radiation damping $\Gamma_R(\Delta_i, t)$. Retaining this last one in expression (30) gives

$$\langle S_y^i(t) \rangle = -\frac{\hbar \omega_0}{4kT} \frac{\chi}{\Omega_i} \sin(\Omega_i t) \exp[-\Gamma_R(\Delta_i, t) t]. \tag{35}$$

The response of the system is given by sum over active spins

$$\langle M_y(t) \rangle = \sum_i \langle S_y^i(t) \rangle = -\frac{g \mu_B \hbar \omega_0 \chi}{4kT} \times \int_{-\infty}^{\infty} n_{ac}(\Delta_i, t) \frac{\sin(\Omega_i t)}{\Omega_i} \exp[-\Gamma_R(\Delta_i, t) t] d\Delta_i, \tag{36}$$

where spectral concentration of the active spins $n_{ac}(\Delta_i, t)$ is given by expression (7).

## 3. Numerical calculations

In experiments [4, 5] power of the excitation pulses expressed in Rabi frequency is varied in range $\chi / 2\pi = 5 \text{ kHz} \div 300 \text{ kHz}$. Figure 1 shows the time evolution of the spectral distribution of the active spins $n_{ac}(\Delta_i, t)$ given by formula (7) for $\chi / 2\pi = 200 \text{ kHz}$. Other parameters involved are related to sample #3 of Ref. 5: $n_0 = 2.4 \times 10^{17} \text{ cm}^{-3}$, $\Delta n_0 \approx 8 \times 10^{15} \text{ cm}^{-3}$, $\sigma / 2\pi = 1 \text{ MHz}$, $T_2 = 75 \text{ μs}$, $T_1 = 1.2 \text{ s}$. All other figures below except Figures 4 and 5 are obtained with use of these parameters.

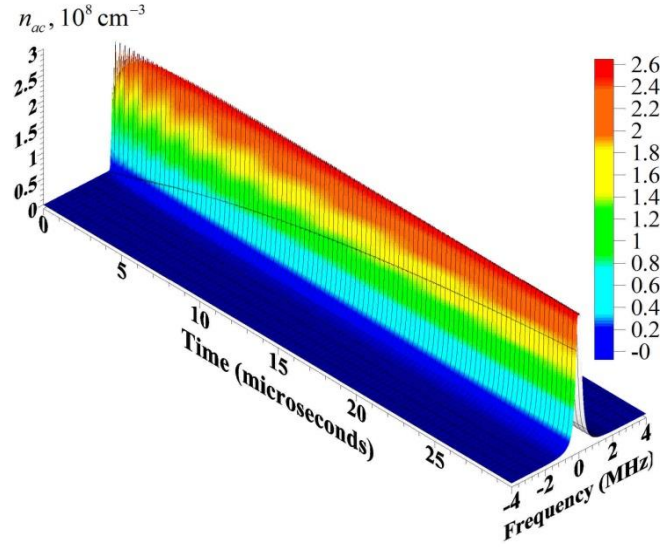

**Fig. 1**. (Color online) Time evolution of the spectral distribution of active spins $n_{ac}(\Delta_i, t)$ (formula (7)) for $\chi/2\pi = 200$ kHz for sample #3 of Ref. 5: $n_0 = 2.4 \times 10^{17}$ cm$^{-3}$, $T_2 = 75$ μs, $T_1 = 1.2$ s, $\sigma/2\pi = 1$ MHz, $T = 4.2$ K.

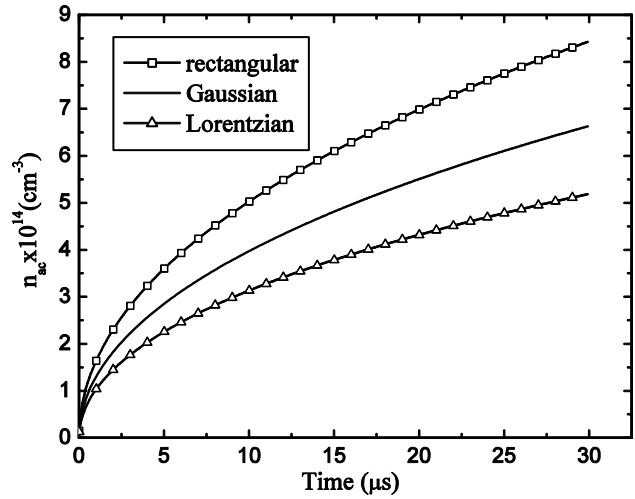

**Fig. 2.** The excitation kinetics of spins, $n_{ac}(t)$, during the nutation process with parameters given in Fig. 1, when inhomogeneous distribution is Gaussian, Lorentzian, or rectangular with one and the same line width.

Time dependence of the concentration of active spins $n_{ac}(t)$ and passive spins $n_{pas}(t)$ are given by formulae (9) and (10), respectively, and $n_{ac}(t)$ is depicted in Fig. 2 with parameters related to sample #3 of Ref. 5. It is obvious from expressions (7) and (9) that the maximally possible value of $n_{ac}(t)$ is equal to $\Delta n_0 / 2$. Figure 2 shows that even with the larger values of the Rabi frequency $(\chi/2\pi = 200$ kHz$)$ only a little part of $n_0$ is participate in the excitation process.

As is seen from expressions (32), (33), and (34), respectively, instead of ordinary constant damping parameter $\Gamma = 1/T_2$, the nutation response is described by intensity-, concentration-, frequency-, and time-dependent parameters $\Gamma_S(\Delta_i, t)$, $\Gamma_I(\Delta_i, t)$, and $\Gamma_R(\Delta_i, t)$, respectively. That is, any i-th spin packet $\Delta_i$ is characterized by its proper time-varying damping parameter.

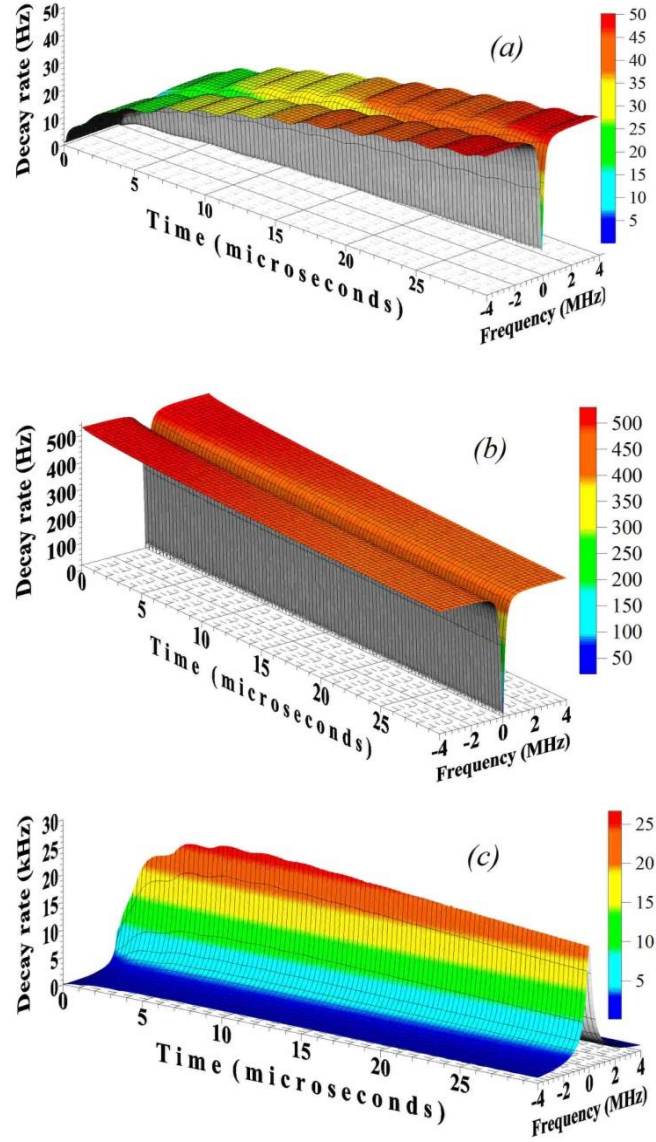

**Fig. 3**. (Color online) Spectral distribution and time dependence of the contributions to the nutation damping parameter $\Gamma(\Delta_i, t)$: (a) from dipolar interactions between active spins, $\Gamma_S(\Delta_i, t)$; (b) from dipolar interactions of active spins with passive ones, $\Gamma_I(\Delta_i, t)$; (c) from radiation damping, $\Gamma_R(\Delta_i, t)$. The used parameters are as in Fig. 1, geometrical factor $a_1 = 2.1\times 10^{-15}\,\text{cm}^3$. Note the scale change along vertical axis.

Figures 3(*a*), 3(*b*), and 3(*c*) depict the $(\Delta_i, t)$ - dependent contributions to the damping from dipolar interactions between active spins, $\Gamma_S(\Delta_i, t)$, of active spins with passive ones, $\Gamma_I(\Delta_i, t)$, and contribution due to the radiation damping, $\Gamma_R(\Delta_i, t)$, respectively (again for $\chi/2\pi = 200$ kHz and with the use of parameters related to sample #3 of Ref. 5). Figure 3(*c*) and what follows are obtained with the value of the geometrical factor $a_1 = 2.1\times 10^{-15}\,\text{cm}^3$. A comparison with Fig. 1 shows that for resonantly excited $(\Delta_i = 0)$ *active* spins contribution $\Gamma_S(\Delta_i, t) \approx 0$ and $\Gamma_I(\Delta_i, t) = 0$. That is, substitution of those in expression (36) instead of $\Gamma_R(\Delta_i, t)$ leads to the negligible damping of nutations.

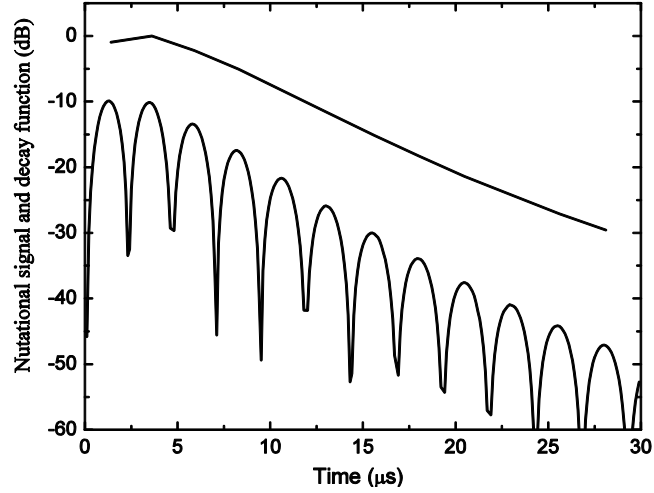

**Fig. 4.** Nutational response and decay function $K(t)$ (top) calculated for sample #2 in Ref. 5 with $\chi/2\pi = 113$ kHz (for other parameters see text).

Figure 4 shows typical nutation response and the decay function $K(t)$ (obtained by comparing maxima of the signal and of the Bessel function $J_0(\chi t)$, see [4, 16]) calculated with the use of parameters $n_0 = 1.6 \times 10^{17}$ cm$^{-3}$, $T_2 = 160$ μs, $T_1 = 0.7$ s related to sample #2 of Ref. 5, for $\chi/2\pi = 113$ kHz (see Fig. 2 in Ref. 5). Note that Heisenberg equations (11) and their solutions are written down without use of the low-power contributions to the damping and hence all the results of the present work are correct for $\Gamma_R \gg \Gamma_0$. From comparison of the nutation response of Fig. 4 with the experimental one in Fig. 2 of Ref. 5 we see that the present theory is correct at the middle part of the nutational process but fails at the beginning (first maximum in Fig. 4), where $\Gamma_R$ is small due to small concentration $n_{ac}(t)$, and at the tail (where the decay function's line is somewhat curved up) where $\Gamma_R$ is small due to decrease of $S_\alpha(t)$.

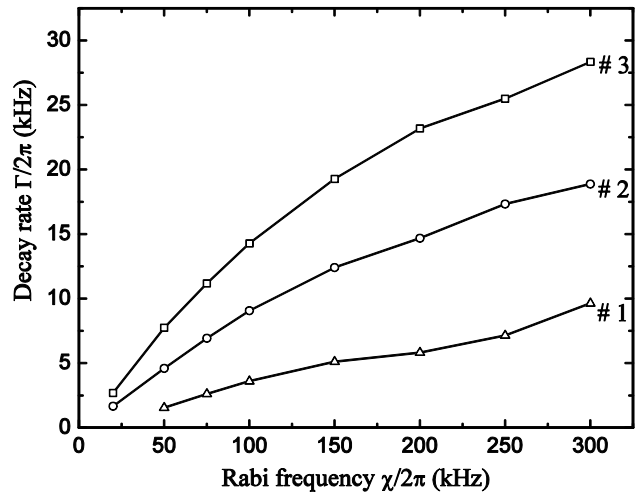

**Fig. 5.** Calculated nutation decay rate $\Gamma$ versus Rabi frequency $\chi$ for three samples of Ref. 5 with concentrations $n_{01} < n_{02} < n_{03}$.

Next Fig. 5 shows the calculated dependence of the nutation decay rate $\Gamma$ on the Rabi frequency $\chi$ for three samples of Ref. 5, obtained by use of $K(t)$ results. Comparison with Fig. 3 of Ref. 5 reveals qualitative coincidence between theory and experiment in the concentration dependence, although the $\chi$-dependence is not so straight line as in the experiment and as should be according to formula (2). If one retains in the figure only data points with $75 \text{ kHz} < \chi/2\pi < 200 \text{ kHz}$ (that is, $\Gamma \gg \Gamma_0$), linear fit gives for $\beta$ in formula (2) values $\beta_1 = 0.025$, $\beta_2 = 0.061$, and $\beta_3 = 0.095$ whereas experiment (Ref. 5) gives $\beta_1 = (4.8 \pm 0.5) \times 10^{-2}$, $\beta_2 = (6.1 \pm 0.5) \times 10^{-2}$, and $\beta_3 = (10.6 \pm 0.5) \times 10^{-2}$ for the three samples, respectively. As it should be, the agreement between theory and experiment is better for sample with larger concentration since $\Gamma \approx \Gamma_R \sim n_{ac} \sim n_0$ and the condition $\Gamma_R \gg \Gamma_0$ is fulfilled easier in this case.

## 4. Discussion

Obtained in preceding sections results can be commented as follows. If the wavelength of the driving field is large compared with the sample dimensions or with the distances between spins, all the spins will interact collectively with a single mode of the field forming the cooperative superradiant Dicke states [18]. As a result, instead of rather weak coupling between a single spin and the electromagnetic field with the coupling strength $g_0$ the strong collective coupling of the ensemble of $N$ spins to the cavity mode with the coupling strength $g_{col} \sim \sqrt{N} g_0$ is realized [19]. For large values of photon number $n_{ph} \gg N$ in the cavity, the strong collective coupling strength $g_{col}$ converges to the classical limit $g_{col} \sim \sqrt{n_{ph}} g_0 \sim \chi$ [19]. In our classical description of the cavity mode the spin-cavity coupling is given by Hamiltonian $H_1^i = \hbar \chi S_x^i$ (13). Since in the whole range of Boscaino experiments [4, 5] condition $\chi \gg \Gamma$ is fulfilled, the strong coupling limit is realized there. The strong coupling leads to the enhancement of the spontaneous emission rate in the resonant cavity [20], and, as a consequence, to the radiation damping. It is remarkable that due to the two-photon excitation procedure in experiments [4, 5] one has a pure spontaneous emission in the cavity.

The inhomogeneous line broadening determines the spectral distribution of the active spins concentration (see Fig. 1) and has drastic influence on the frequency dependence of the decay rates due to the dephasing of the spins during the time evolution (Fig. 3). As is seen from expressions (25), all the three decay mechanisms $\xi_i^{S,I,R}$ affect only the oscillating parts of the spin components. Further, the mechanisms affect the longitudinal and transverse spin components equally, serving as phase and energy relaxation mechanisms simultaneously.

It was shown [21,22,23] that the decay rate of the Rabi oscillations depends not only on the width of the spectral distribution but on its form as well. In particular, the decay rate is significantly slower in the case of the Gauss distribution (the so-called cavity protection effect, Ref. 22) than for the Lorentzian one. To test this effect in our case, we compare the nutation decay rates for Gaussian, Lorentzian, and rectangular forms of the inhomogeneous broadening *with one and the same value of the line width*. It is not so easy to observe the differences, however, between 3D pictures of $\Gamma_R(\Delta_i, t)$ (similar to Fig. 3(c)) for these three forms.

Hence, in Fig. 6 we compare the resonance components $\Gamma_R(\Delta_i = 0, t)$ (i.e. the 'hump' line in Fig. 3(c)). Note that the similar relationship is retained for $\Gamma_R(\Delta_i \neq 0, t)$ as well. As is seen from formula (34), $\Gamma_R(\Delta_i, t)$ is given by product of the concentration $n_{ac}(t)$ and integral $I(\Delta_i, t) = a_1 \int d\Delta_j f(\Delta_j) |\gamma_{ij}^0|$. It is clear from definitions of the rectangular, Gaussian, and Lorentzian distribution functions that $n_{ac}^r(\Delta, t) > n_{ac}^G(\Delta, t) > n_{ac}^L(\Delta, t)$ excepting far wings (with insufficient contribution to $n_{ac}(t)$ due to small $s(\Delta)$ there, see formulas (6) and (7)) and the same inequality is true for $n_{ac}(t)$ as is shown in Fig. 2. On the other hand, the integral pictured for $\Delta_i = 0$ in Fig. 7 is initially less in the Lorentzian case but gradually becomes larger than that for the rectangular and Gaussian distributions.

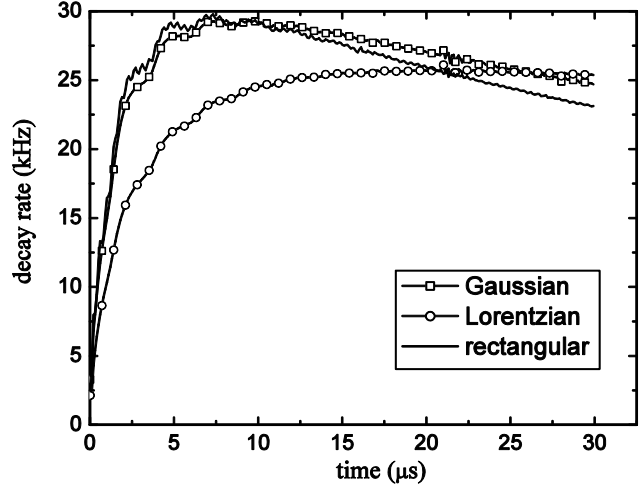

**Fig. 6**. Time dependence of decay rate $\Gamma_R(\Delta_i = 0, t)/2\pi$ of resonantly excited $(\Delta_i = 0)$ spins when inhomogeneous distribution function $f(\Delta_j)$ is of a Gaussian, Lorentzian, or rectangular type.

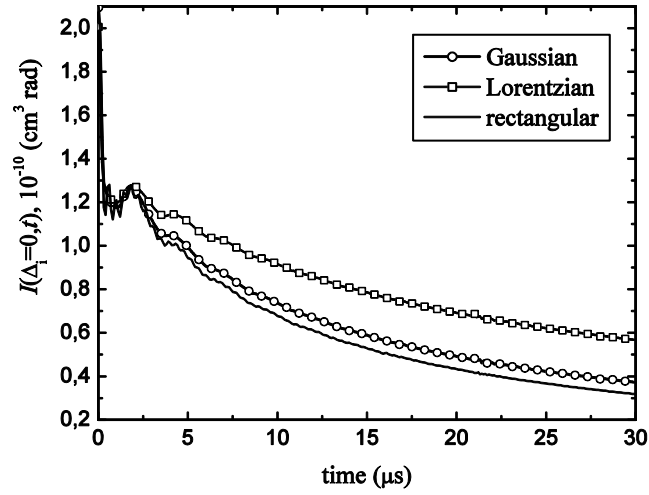

**Fig. 7**. Time dependence of the integral $I(\Delta_i = 0, t) = a_i \int d\Delta_j f(\Delta_j) |\gamma_{ij}^0|$ when inhomogeneous distribution function $f(\Delta_j)$ is of a Gaussian, Lorentzian or rectangular type.

Again, a such behavior is retained for $\Delta_i \neq 0$. As a result, in transient nutation regime the decay rate $\Gamma_R(\Delta_i, t)$ and decay function $K(t) = \exp(-\Gamma t)$ are more intensive in the case of the Lorentzian distribution, as is seen from Fig. 8 for the decay function $K(t)$. Here, after removal of first three data points, a linear fit gives for the overall decay rate $\Gamma$ numerical values $\Gamma_r/2\pi = 20.75$ (kHz), $\Gamma_G/2\pi = 22.96$ (kHz) and $\Gamma_L/2\pi = 25.8$ (kHz).

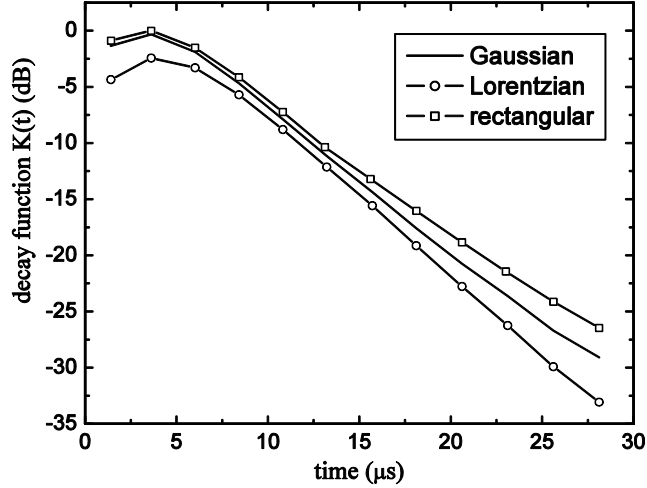

**Fig. 8**. Decay function $K(t) = \exp(-\Gamma t)$ when inhomogeneous distribution function $f(\Delta_j)$ is of a Gaussian, Lorentzian, or rectangular type.

## 5. Conclusions

In conclusion, we have considered driving-field intensity and concentration dependent damping of nutation signals observed in various diluted paramagnetic solids. The main results of the paper can be summarized as follows. 1) We solved correctly formulated kinetic equation for the spectral distribution $n_{ac}(\Delta,t)$ and found full concentration $n_{ac}(t)$ of the active spins excited during the excitation pulse. The distribution is of crucial importance for the damping mechanisms and for the nutation response in general. 2) We give correct expression for the equilibrium density matrix $\rho(t)$ which is time dependent because of time dependence of $n_{ac}(t)$. 3) We have elaborated damping mechanisms due to the dipole-dipole interaction between active spins and due to interaction of active spins with unexcited (passive) ones with concentration $n_{pas}(t)$. It is shown that in inhomogeneously broadened spin system the dipole interactions of both type give negligible contribution to the nutations decay. 4) A radiation damping mechanism is elaborated well describing the experimentally observed properties of nutations in diluted paramagnetic solids. It is shown that instead of usual Bloch decay parameters $T_1$ and $T_2$, the damping caused by this mechanism is described by the intensity-, concentration-, frequency- and time-dependent distribution of decay parameter $\Gamma_R(\Delta,t)$ and the dependencies are displayed. The considered mechanism affects the longitudinal and transverse spin components equally, serving as phase and energy relaxation mechanism simultaneously. 5) It is shown that the decay rate of the Rabi oscillations depends not only on the width of the inhomogeneous spectral distribution but on its form as well, in good accordance with the so-called cavity protection effect.

**Acknowledgment**

The authors thank prof. B. Z. Malkin for useful discussions and valuable comments.